\documentclass[doublespacing]{elsart}

\usepackage{icarus}

\usepackage{natbib}

\bibpunct{(}{)}{;}{a}{,}{,}

\usepackage{graphicx}

\def\kg{\,\mathrm{kg}}
\def\km{\,\mathrm{km}}
\def\ergps{\,\mathrm{erg\, s^{-1}}}

\def\K{\,\mathrm{K}}

\def\MS{M_\mathrm{S}}
\def\MT{M_\mathrm{T}}

\def\Phie{\Phi_\mathrm{e}}
\def\Phieself{\Phi_\mathrm{e}^\mathrm{self}}

\def\kice{k_\mathrm{ice}}
\def\cp{c_\mathrm{p}}
\def\DT{\Delta T}
\def\loss{\dot E_\mathrm{loss}}

\def\heateq{\dot E_\mathrm{heat}^\mathrm{eq}}

\def\Eelaseq{E_\mathrm{elas}^\mathrm{eq}}

\def\rhoi{\rho_\mathrm{i1}}
\def\rhoo{\rho_\mathrm{o}}
\def\rhohp{\rho_\mathrm{hp}}
\def\rhoc{\rho_\mathrm{c}}
\def\rhoave{\rho_\mathrm{ave}}
\def\mui{\mu_\mathrm{i1}}
\def\muo{\mu_\mathrm{o}}
\def\muhp{\mu_\mathrm{hp}}
\def\muc{\mu_\mathrm{c}}
\def\Ro{R_\mathrm{o}}
\def\Rhp{R_\mathrm{hp}}
\def\Rc{R_\mathrm{c}}

\def\ktwo{k_\mathrm{2}}
\def\kteq{k_\mathrm{2,eq}}
\def\kteqmax{k_\mathrm{2,eq}^\mathrm{max}}
\def\kf{k_\mathrm{f}}

\def\ksym{k_\mathrm{2,0}}
\def\kpro{k_\mathrm{2,2}}
\def\kretro{k_\mathrm{2,-2}}
\def\Phiesym{\Phi_\mathrm{e}^\mathrm{2,0}}
\def\Phiepro{\Phi_\mathrm{e}^\mathrm{2,2}}
\def\Phieretro{\Phi_\mathrm{e}^\mathrm{2,-2}}

\def\Ptt{P_2^2}
\def\Ptz{P_2^0}

\def\taue{\tau_e}

\def\enow{e_\mathrm{now}}

\def\ktobs{k_\mathrm{2,obs}}
\def\ktg{k_\mathrm{2,g}}
\def\wdyn{\omega_\mathrm{dyn}}
\def\nmode{n_\mathrm{mode}}
\def\norb{n_\mathrm{orb}}

\def\wclose{\omega_\alpha}
\def\dw{\delta\omega}

\def\Emode{E_\mathrm{mode}}
\def\xir{\xi_r}
\def\xireq{\xi_r^\mathrm{eq}}
\def\xih{\xi_\perp}
\def\xiheq{\xi_\perp^\mathrm{eq}}
\def\xirsurf{\xi_r^\mathrm{surf}}

\def\xihsurf{\xi_\perp^\mathrm{surf}}

\def\kr{k_r}
\def\kh{k_\perp}

\def\ttra{t_\mathrm{tra}}
\def\tcoh{t_\mathrm{coh}}
\def\Dw{\Delta\omega}
\def\OT{\Omega_\mathrm{T}}

\def\Cassini{{\it Cassini}\,\,}

\def\Cassini{{\it Cassini}\,\,}

\begin{document}

\begin{frontmatter}



\title{Titan's Dynamic Love Number Implies Stably-Stratified Ocean}


\author[luan]{Jing Luan}, 

\address[luan]{Institute for Advanced Study, 
               Princeton, NJ 08540 (U.S.A.)}
             
\begin{center}
\scriptsize
Copyright \copyright\ 2019 Jing Luan
\end{center}


%
%
%
%
%


\end{frontmatter}



\begin{flushleft}
\vspace{1cm}
Number of pages: \pageref{lastpage} \\
Number of tables: \ref{lasttable}\\
Number of figures: \ref{lastfig}\\
\end{flushleft}


\begin{pagetwo}{Stably Stratified Ocean in Titan}

Jing Luan \\
Institute for Advanced Study\\
BH-128, 1 Einstein Drive\\
Princeton, NJ 08540, USA. \\
\\
Email: jingluan@ias.edu\\
Phone: (626) 437-6258 

\end{pagetwo}

\begin{abstract}

The dynamic quadrupole Love number of Titan measured by \Cassini is $\ktobs=0.616\pm 0.067$, strongly indicating a global subsurface ocean. However, the theoretical Love number due to equilibrium tides is at most $\kteqmax\approx 0.48$ in the absence of an ice shell on top of the ocean. In reality, there is an outer ice shell of thickness $\sim 100\km$, reducing the equilibrium-tide Love number to $\kteq\approx 0.42$. Therefore, other types of tidal response, like dynamic tides, may be also present in Titan.

We propose that the ocean is stably stratified. As a result, there exist standing ocean waves (gravity modes) with eigen-frequencies close to the tidal frequency.  Such a gravity mode (g-mode) is resonantly excited. It bends the outer ice shell radially and thus enhances the dynamic Love number by $\ktg$. In order for $\ktg$ to account for the discrepancy between $\kteq$ and $\ktobs$, the Brunt-Vaisala frequency in the ocean is required to be $3.3\times 10^{-4}\,\mathrm{rad\, s^{-1}}$. It is compatible with the volatile-rich model for Titan that was proposed to explain the methane-rich atmosphere.

The three components of the tidal potential with azimuthal degrees, $m=-2,0,2$, correspond to the three components of the quadrupole Love number, $\kretro$, $\ksym$ and $\kpro$. They can excite retrograde, axisymmetric and prograde g-modes equally in the absence of rotation. However, Coriolis force induced by Titan's rotation breaks the symmetry among these modes. Most likely, only one of the Love-number components is significantly enhanced by a g-mode, while the other two are still attributed to equilibrium tides. This prediction is testable by observation. If confirmed, the smaller components of the Love number can be used to constrain the thickness of the outer ice shell.

\end{abstract}



\section{Introduction}
Titan, the most massive satellite of Saturn, is on a $15.9$-day orbit with eccentricity, $e\approx 0.0288$, and inclination, $I\approx 0.35^\circ$, with respect to the equatorial plane of Saturn \citep{Ephemeris}. Titan's spin frequency, $\OT$, is synchronized with its orbital mean motion, $\norb$. Saturn exerts a tidal potential on Titan. Its leading order is quadrupole, $\Phi_\mathrm{tide}$. It contains a static or time-averaged part, $\Phi_\mathrm{static}$.  $\Phi_\mathrm{tide}$ reduces to $\Phi_\mathrm{static}$ for a synchronously rotating satellite on a circular orbit. The tidal deformation induced by $\Phi_\mathrm{static}$ has relaxed to that of a fluid body. The tidal deformation yields a gravitational potential that is proportional to $\Phi_\mathrm{static}$ by the dimensionless fluid Love number, $\kf$. Titan's $\kf$ is about $1.01$, measured by \cite{Iess-2010} and recently confirmed by \cite{Durante}. The fluid Love number for a homogeneous body is $1.5$. Mass concentrating towards the center results in $\kf<1.5$. The fluid Love number and hydrostatic gravitational parameters \citep{Durante} constrain the mass distribution in Titan \citep[e.g.][]{Rappaport,Baland-2014}.

 For a synchronously rotating satellite on an eccentric orbit, the distance and the longitude of Saturn in the rest frame of the satellite both oscillate by a fraction $\sim e$ at frequency $\norb$. Consequently, $\Phi_\mathrm{tide}$ contains a time-varying part, $\Phie$. The leading order of $\Phie$ is $\propto e$ and oscillates with frequency $\norb$ (Appendix~\ref{app:love-number-3-components}). The corresponding tidal deformation depends on material rigidity of the solid parts of Titan, because the elastic response cannot be relaxed within one orbital period. The induced gravitational potential is proportional to $\Phie$ by the dynamic quadrupole Love number, $\ktwo$. We mainly deal with $\ktwo$ and call it the dynamic Love number. We refer to $\kf$ as the fluid Love number. For a coherent solid body, $\ktwo\ll\kf$. A global liquid layer raises $\ktwo$ towards $\kf$, but still $\ktwo<\kf$ unless the whole body is liquid. The gravitational potential of Titan was measured by tracking Doppler shifted radio signals from \Cassini during its flybys of Titan. The static tidal response is associated with $\kf\Phi_\mathrm{static}$. The dynamic tidal response is associated with $\ktwo\Phie\sim \ktwo e\Phi_\mathrm{static}$. Because $\e\ll 1$ and $\ktwo<\kf$, it is more difficult to measure $\ktwo$ than $\kf$. \cite{Iess-2012} measured Titan's dynamic quadrupole Love number and recently \cite{Durante} improved it to $\ktobs=0.616\pm 0.067\, (1\sigma)$.\footnote{\cite{Durante} also tried to fit the time lag between $\Phie$ and $\ktwo\Phie$. It is consistent with zero and its upper limit is about a tenth of the orbital period. The imaginary part of $\ktwo$ is $k_\mathrm{2}^{im}\approx 0.082\pm 0.118$, which is poorly constrained by observation. We only discuss the real part of $\ktwo$ in this paper.} The $\ktobs$ is so close to $\kf$ that it strongly indicates an internal ocean overlaid by a thin outer ice shell. 

The tidal response of Titan is usually treated as equilibrium tides because the dynamic frequency $\wdyn=(G\MT/R^3)^{1/2}\approx 7.25\times 10^{-4}\,\mathrm{rad\, s^{-1}}$, is much greater than the tidal frequency, $\norb\approx 4.56\times 10^{-6}\,\mathrm{rad\, s^{-1}}$. The dynamic Love number attributed to equilibrium tides is $\kteq$. It increases with a thinner outer ice shell, because the shell resists tidal deformation. It reaches the maximum value $\kteqmax\approx 0.48$ in the absence of an outer ice shell, but $\kteqmax$ is still more than $1\sigma$ below $\ktobs$. The theoretical $\kteqmax$ mainly depends on the mass and size of the solid core and the rigidity of the core. The former is well constrained by the observed static gravitational potential of Titan \citep{Iess-2010, Durante}. The latter may be artificially reduced in order to increase the theoretical $\kteqmax$. However, as already pointed out in \cite{Rappaport} and \cite{Durante}, one has to adopt some yet unknown peculiar elastic-viscous model for the core or to reconcile the tension between $\kteqmax$ and $\ktobs$. Another way to reconcile the tension is to assume dense water \citep{Baland-2014}. We constrain our discussion to normal values for physical quantities as shown in Table~\ref{tab:parameters}. 

In reality, the outer ice shell has a finite thickness ($d$) and thus makes $\kteq$ smaller than $\kteqmax$. Considering the balance between the radiogenic heating of the rocky core and the conductive heat loss rate through the outer ice shell, we estimate $d\sim 100\km$.\footnote{We adopt the specific radiogenic heating rate of the Earth's mantle, $7.4\times 10^{-8}\,\mathrm{erg\, s^{-1}\, g^{-1}}$. The intrinsic heating of the Earth is partly due to radiogenic heating and primordial heating. The two resources are believed to be comparable. Given that Titan's rocky core is of half of its total mass, $M_c\sim 0.5\MT\sim 6.7\times 10^{25}\,\mathrm{g}$, we obtain a heating rate $\sim 5\times 10^{18}\ergps$. It could balance the conductive heat loss rate, $\loss\sim 4\pi R^2 \kice \DT/d\sim 3.0\times 10^{18}(d/100\km)^{-1}\ergps$, at ice shell thickness $d\sim 100\km$.}\footnote{The obliquity of Titan, the angle between the spin axis and the orbital normal, was measured to be $0.3^\circ$ \citep{Meriggiola}. Because it is bigger than $0.12^\circ$ expected for a completely solid Titan in a Cassini state, \cite{Bills} suggested that there existed a subsurface ocean. Unfortunately, the obliquity cannot put a tight constraint on $d$. According to \cite{Baland}, $d$ could range from a few tens of kilometers to two hundred kilometers.} With $d=100\km$, $\kteq\approx 0.42$, according to the tidal model described in Appendix~\ref{app:eq-tides}. This value is about 3$\sigma$ below $\ktobs$. One might ask whether tidal heating could generate enough power to maintain a shell much thinner than $100\km$ and thus boost up $\kteq$ to almost its maximum value. We checked several mechanisms for tidal heating but none of them can do so (Appendix~\ref{app:tidal-heating}).


Therefore, we are in need of a mechanism that could enhance the dynamic Love number by $(\ktobs-\kteq)/\kteq\approx 46\%$. Since $\ktwo$ is mainly attributed to the radial displacement of Titan's surface,\footnote{At Titan's surface, the density jumps from that of ice to almost zero, therefore the radial displacement of Titan's surface leads to column mass excess (or deficit) of order $\rhoi\xirsurf$. This column mass excess (deficit) yields a perturbation to the gravitational potential of Titan. In Titan's interior, e.g. at the interface between the rocky core and the high-pressure ice layer, the density changes by $\rhoc-\rhohp$. The radial displacement at this interface also leads to a perturbation to the gravitational potential of Titan. All these perturbations add up to the total induced tidal potential $\ktwo\Phie$. Because the density jump at Titan's surface is the biggest, $\ktwo$ is mainly attributed to the radial displacement at Titan's surface.} increasing $\ktwo$ requires enlarging the surface's radial displacement. We propose that resonantly excited gravity modes (g-mode) in the presumably stably-stratified ocean could account for the enhancement. A g-mode is a standing ocean wave which contains multiple half-wavelengths in the radial direction \citep{Unno}. The number of nodes along the radial direction, $\nmode$, is known as the radial order. Stable stratification slows down the group velocity of ocean waves since $v_g=\partial\omega/\partial\kr\sim \omega^2/(N\kh)$, where $\kh=\sqrt{\Lambda}/r$ is the horizontal wavenumber and $\Lambda=\ell(\ell+1)=6$ for quadrupole modes \citep{Unno}. Here $N>0$ is the Brunt-Vaisala frequency. A larger $N$ means the ocean is more stably-stratified and thus leads to a slower group velocity. In general, a stabilizing composition gradient or temperature gradient can both lead to a positive $N$. The Earth ocean is stably stratified mainly by a stabilizing temperature gradient. Thanks to the slow group velocity, g-modes have eigen-frequencies well below the dynamic frequency, $\wdyn$. Thus it is possible to find a high-order g-mode whose eigen-frequency matches the tidal frequency, $\norb$. Such a mode could be resonantly excited to a large enough amplitude to account for the enhancement in the dynamic Love number. 

A g-mode acts like a damped harmonic oscillator driven near resonance with the frequency mismatch much bigger than the linear damping rate. Its amplitude is inversely proportional to the frequency mismatch between the eigen-frequency and the driving (tidal) frequency, $\dw\equiv \omega-\norb$. The tidal frequency generally falls between the eigen-frequencies of two g-modes with neighboring radial orders. Therefore,  $\dw$ is generally less than half of the frequency separation, $\Dw$, between the two modes. The frequency separation is mainly determined by the Brunt-Vaisala frequency. Assuming the general case, $\dw\sim 0.5\Dw$, we estimate $N_\mathrm{req}\sim 3.3\times 10^{-4}\,\mathrm{rad\, s^{-1}}$ in order to enhance the theoretical $\ktwo$ to match what's observed. The required value for $N$ is compatible with the volatile-rich model for Titan \citep{Lunine-Stevenson}. In this model, the outer ice shell is enriched with methane, while the ice layer beneath the ocean is enriched with ammonia. Since the molecular weight of ammonia is greater than that of methane, there may exist a stabilizing composition gradient in the ocean yielding $N\sim 3.6\times 10^{-4}\,\mathrm{rad\, s^{-1}}$. 

Our proposal is probably testable with the current \Cassini data on gravity measurements of Titan. The quadrupole Love number has three even azimuthal degrees, $m=-2,0,2$, corresponding to three components, $\kretro,\ksym,\kpro$. Equilibrium tides are equally excited at these three azimuthal degrees. However, g-modes are modified by Titan's rotation and thus cannot be equally excited at $m=-2,0,2$. Therefore, the three components of the Love number should be different. We recommend fixing two components of the Love number at the theoretical $\kteq$ and fitting the third component. If this algorithm improves the fitting significantly, it would be a great support for our proposal. The theoretical $\kteq$ is a function of the thickness of the outer ice shell (Figure~\ref{fig:k2-Heat-eqtides}). Therefore this procedure may also provide a constraint on $d$. If our proposal passes this test, then it is worth carrying out further tests which we will discuss later.

This introduction section already presents our major results. Readers interested in more detailed calculations for the amplitude of g-modes and the Brunt-Vaisala frequency are directed to the next section. Observers interested in testing our proposal are directed to the final section. Readers interested in tidal heating mechanisms are directed to  Appendix~\ref{app:tidal-heating}. We discuss scenarios involving deep ocean standing waves (g-modes), shallow ocean waves and inertial wave attractors. None of them could produce enough tidal heating to keep the outer ice shell much thinner than $\sim 100\km$ in order to enhance the theoretical $\kteq$. However, they apply broadly to satellites of similar size to Titan which contain internal oceans. 

\section{Heuristic Treatment of Gravity Modes in Titan's Ocean}\label{sec:Love-number}
\cite{Goodman-Dickson} studied gravity waves tidally excited in late-type stars. The physical situation is similar to Titan. The outer ice shell of Titan is analogous to the outer convective zone in the star. The presumably stably-stratified ocean is akin to the inner radiative zone in the star. According to \cite{Goodman-Dickson}, at the outer boundary of the stably-stratified region, the wave's horizontal displacement is similar to that due to equilibrium tides,   
\begin{equation}\label{eq:wave-eq}
    \xih^\mathrm{wave}\sim \xih^\mathrm{eq}\, , \,\, (\mathrm{at}\,\, r=R-d)\, ,
\end{equation}
if the wave does not reflect back to the outer boundary of the stably-stratified region. For an incompressible ocean of depth $H$, the equilibrium tides satisfy $\xireq/H\sim \kh\xiheq$, and therefore we have
\begin{equation}\label{eq:eqr-eqh}
    \xiheq\sim {\xireq\over H\kh}\sim \xireq{R\over \sqrt{\Lambda}H}\, .
\end{equation}

We assume that the bottom of Titan's ocean is smooth enough so that the wave reflects at the inner boundary and travels back to where it was excited. The oscillatory velocity associated with the wave is still in phase with the tidal force, as long as the travel time, $\ttra$, is shorter than the coherent time, $\tcoh$. In this case, the amplitude of the horizontal displacment grows by an additional $\xih^\mathrm{wave}$.

The travel time is the time for the wave package to travel through the ocean radially,
\begin{equation}\label{eq:ttra}
    \ttra\sim\int_\mathrm{ocean} {dr\over v_g}\sim {1\over\omega}\int_\mathrm{ocean}{dr\kr}\sim {\nmode\over\omega}\sim {1\over \Dw}\, ,
\end{equation}
where the radial order, $\nmode$, is the number of half-wavelengths along the radial direction \citep{Unno},
\begin{equation}\label{eq:nmode}
    \nmode \sim {1\over\pi}\int_\mathrm{ocean} dr\kr\sim{1\over\pi}\int_\mathrm{ocean}dr \kh{N\over\omega} \sim {\sqrt{\Lambda}H\over\pi R}{N\over\omega}\, . 
\end{equation}
The frequency separation of g-modes with neighboring radial orders is $\Dw\equiv \omega_\mathrm{\nmode}-\omega_\mathrm{\nmode+1}\sim \omega/\nmode$. We adopt the interior model in \cite{Rappaport}. From inside out, Titan contains an inner rocky core, a high-pressure ice layer, an ocean layer, and an outer ice shell. All parameters are listed in Table~\ref{tab:parameters}. The bottom of the ocean is at $\Rhp\approx 2275\km$. Given Titan's radius $R\approx 2575\km$ and $d\sim 100\km$, we have $H=R-d-\Rhp\sim 200\km$. 

With $N\gg\norb\sim\omega$, $\nmode$ is much greater than unity and thus $\Dw\sim\omega/\nmode$ is much smaller than $\omega$, which means the spectrum of g-modes is dense near the tidal frequency. Therefore, the chance is high to find a g-mode (corresponding to an integer $\nmode=\alpha$) with eigen-frequency $\wclose\approx\norb$ . The frequency mismatch $\dw\equiv\wclose-\norb$ typically falls in the range $-0.5\Dw<\dw<0.5\Dw$. In order for the g-mode to enhance $\ktwo$, the radial displacement of the g-mode must be in phase with the tidal potential. This requires $\dw>0$. The chance for a positive $\dw$ is $50\%$ if $\norb$ locates randomly in between two neighboring eigen-frequencies. We do not see a physical reason why a positive $\dw$ should be chosen by Titan. In the case of $\dw<0$, the g-mode should reduce $\ktwo$. We admit this is a caveat for our proposal. Since the current time is probably not a special one in the evolution history of Titan, $\norb$ is not particularly close to any mode's eigen-frequency. Therefore, we choose $\dw\sim 0.5\Dw$.

The frequency mismatch would `drag' the mode out of phase with the tidal force after a time $\sim\pi/\dw$. Damping or scattering of the wave also causes a phase difference of order $\pi$ after time $\sim\pi/\gamma$, where $\gamma$ is the linear damping or scattering rate. Nonlinear damping is probably negligible because the tides are well within the linear regime, $\xih\kh\ll 1$. Therefore, the coherent time, $\tcoh$, defined as the time it takes for the phase difference to reach $\pi$ is
\begin{equation}\label{eq:tcoh}
    \tcoh\sim {\pi\over (\dw^2+\gamma^2)^{1/2}}\sim {\pi\over\dw}\, ,
\end{equation}
where we have assumed $\gamma\ll \dw$. As long as the boundaries of the ocean layer are reasonably smooth, this is a good assumption.  

Within $\tcoh$, the horizontal displacement of the resonant g-mode increases by $\xih^\mathrm{wave}$ every travel time, $\ttra$. After $\tcoh$, the horizontal displacement reaches the saturation value,
\begin{equation}\label{eq:mode-wave}
    \xih^\mathrm{mode} \sim \xih^\mathrm{wave}{\tcoh\over\ttra}\, .
\end{equation}
After a few multiples of $\pi/\gamma$, the g-mode's amplitude is stabilized at this saturation level. One could understand the statement above using a damped harmonic oscillator driven near resonance with $\dw\gg\gamma$. 

Combining Equations~(\ref{eq:wave-eq}), (\ref{eq:eqr-eqh}) and (\ref{eq:mode-wave}), we obtain
\begin{equation}
    \xih^\mathrm{mode}\sim {\pi R\over \sqrt{\Lambda} H}{\Dw\over\dw} \left.{\xireq}\right|_\mathrm{surf}\, .
\end{equation}
In the equation above, we should use $\xireq$ at $r=R-d$. However, with $d\ll R$, $\xireq$ hardly changes across the ice shell. Thus we adopt the radial displacement of Titan's surface due to equilibrium tides, denoted by $\left.{\xireq}\right|_\mathrm{surf}$, which is about $ 460\,\mathrm{cm}$. 

Gravity modes are almost incompressible, satisfying $\kh\xih\sim\kr\xir$ and $\kh/\kr \sim \omega/N$ \citep{Unno}. According to Equation~(\ref{eq:nmode}), $\omega/N\sim {\sqrt{\Lambda}H\over\pi R}{\Dw/\omega}$. Thus the radial displacement of the g-mode is
\begin{equation}\label{eq:mode-eq-r}
    \xir^\mathrm{mode}\sim {\Dw^2\over\omega\,\dw} \left.{\xireq}\right|_\mathrm{surf}\, .
\end{equation}
The g-mode forces the outer ice shell to move radially by almost the same amount as demonstrated in Appendix~\ref{app:xirsurf-xiro-relation}. The ice shell's rigidity is so weak compared to Titan's gravity that the shell bends as much as the underlying ocean `wants' it to. 

In order to enhance the theoretical Love number by $(\ktobs-\kteq)/\kteq\sim 46\%$, we need the total surface radial displacement to be $1.46\left.{\xireq}\right|_\mathrm{surf}$. Thus $\xir^\mathrm{mode}$ needs to be $0.46\left.{\xireq}\right|_\mathrm{surf}$. Adopting $\xir^\mathrm{mode}\sim 0.46 \left.{\xireq}\right|_\mathrm{surf}$ and $\dw\sim 0.5\Dw$, Equation~(\ref{eq:mode-eq-r}) yields $\Dw\sim 0.23\omega$. Combining $\nmode\sim\omega/\Dw$ and Equation~(\ref{eq:nmode}), we obtain
\begin{equation}
    N_\mathrm{req}\sim \norb{\omega\over\Dw}{\pi R\over\sqrt{\Lambda} H}\sim 3.3\times 10^{-4}\,\mathrm{rad\, s^{-1}}\, ,
\end{equation}
where `req' means this is the Brunt-Vaisala frequency `required' to increase the theoretical $\ktwo$ by the right amount. 

Next, we speculate what physical reason might cause stable stratification in Titan's ocean. \cite{Lunine-Stevenson} proposed a volatile-rich model for Titan. Their model can refill Titan's atmosphere with methane which otherwise should have been depleted by photo-chemical reactions. They proposed that Titan formed by accreting rock and ice in the circum-Saturn nebula. In the nebula, all the $\mathrm{H_2O}$ may be combined with volatile molecules as either Ammonia Monohydrate $\mathrm{NH_3.H_2O}$ or Clathrate $\mathrm{CH_4.6H_2O}$. The volatile content of Titan evolves with time \citep{Tobie}. Since Titan's atmosphere is rich in methane now, we speculate the outer ice shell is still composed of Clathrate. The ice in the deep interior may still remain as Ammonia Monohydrate.  $\mathrm{NH_3.H_2O}$ is denser than $\mathrm{CH_4.6H_2O}$ by $\Delta\rho\sim 0.02\,\mathrm{g\, cm^{-3}}$ \citep{Stevenson}. There might be a stabilizing composition gradient in the ocean, leading to
\begin{equation}
    N\sim \left(-{g\over \rho}\left.{d\rho\over dr}\right|_\mathrm{comp}\right)^{1\over 2} \sim \left(-{g\over \rho}{\Delta\rho\over H}\right)^{1\over 2}\sim 3.6\times 10^{-4}\,\mathrm{rad\, s^{-1}}\, ,\label{eq:N}
\end{equation}
where we employ $g\approx 135\,\mathrm{cm\, s^{-2}}$ and $\rho\approx 1\,\mathrm{g\, cm^{-3}}$. This Brunt-Vaisala frequency is similar to $N_\mathrm{req}$, which may be a coincidence, but at least $N_\mathrm{req}$ is not an excessively unreasonable value. Questions like whether such a composition gradient is chemically favorable in the ocean, whether it may be destroyed by convection, and whether there is a salinity gradient or a temperature gradient as well ask for further investigations. Some of them are difficult to address because ammonia hydrates under high pressure are poorly understood due to the lack of experimental data \citep[e.g.][]{Grasset}.

\section{Discussion}\label{sec:summary}
We discuss predictions testable by observations in this section. All the calculations above neglect Titan's rotation. In reality, Titan is synchronously rotating, $\Omega_T=\norb$. Therefore, the quadrupole tidal potential can be separated into three components with azimuthal degrees $m=-2,0,2$ respectively. They have negative, zero and positive azimuthal phase speeds in the rest frame of Titan. We call them retrograde, axisymmetric and prograde components. The tidal responses to them can be described by the retrograde, axisymmetric and prograde Love numbers, denoted by $\kretro, \ksym, \kpro$. They are defined in Appendix~\ref{app:love-number-3-components}. 

Coriolis forces induced by rotation modify the angular patterns of retrograde, axisymmetric and prograde g-modes differently \citep{Townsend}. The difference is significant, because $\OT\sim\omega$. Therefore, retrograde, axisymmetric and prograde g-modes are not equally excited. Consequently, their enhancements to the corresponding components of $\ktwo$ are different. On the other hand, equilibrium tides contribute to $\kretro$, $\ksym$ and $\kpro$ equally, because $\OT\ll\wdyn$. For example, an axisymmetric mode may be preferentially excited, while its prograde and retrograde counterparts are not. As a result, $\ksym$ is enhanced by a mode, whereas $\kpro$ and $\kretro$ are still only due to equilibrium tides. In this case, we expect $\ksym\sim 0.6$ and $\kpro\sim\kretro\sim \kteq\sim 0.42$. We examine the fitting formulas adopted by \cite{Iess-2012} and \cite{Durante} and realize that they assume $\ksym=\kpro=\kretro=\ktwo$. It is true for equilibrium tides and thus is a natural assumption. Applying fitting formulas distinguishing the three components of the dynamic Love number would test our proposal. We recommend adopting $\kteq\approx 0.42$ for two components of $\ktwo$ and fitting the third one using the gravity data. If this algorithm improves the fitting significantly, our proposal would look promising. It is ideal to fit the three components simultaneously. The largest component may be enhanced by a g-mode, while the two smaller ones may be the same and thus are due to equilibrium tides only. Then the latter could be used to indicate the ice shell thickness using the upper panel of Figure~\ref{fig:k2-Heat-eqtides}.

Moreover, the angular pattern of g-modes along the latitude in a rotating frame is described by Hough functions instead of associated Legendre polynomials \citep{Townsend}. Towards large value of $\nu\equiv |2\OT/\omega|$, the Hough functions for g-modes remain close to zero at high latitude with $|\cos\theta|\geq |\nu|^{-1}$, where $\theta$ is the colatitude.\footnote{There is another set of Hough functions concentrating near the poles. They correspond to r-modes, whose eigen-frequencies are not resonant with $\norb$ \citep[Fig.1 in ][]{Townsend}. Therefore we do not discuss them.} For Titan, $\nu=2$, the Hough functions concentrate within latitudes lower than $30^\circ$. The 110th flyby of Titan, labeled by T110 in \cite{Durante}, occurred at a high latitude, $74.8^\circ\mathrm{N}$. There are three flybys, T022, T045 and T068, occurring at latitudes greater than $30^\circ$, respectively $45.4^\circ\mathrm{N}$, $43.5^\circ\mathrm{S}$ and $48.9^\circ\mathrm{S}$. These flybys are particularly valuable in distinguishing Hough functions from associated Legendre polynomials.

Because Hough functions are not orthogonal associated Legendre polynomials, a Hough function with $|m|=2$ usually has non-negligible projections onto associated Legendre polynomials with $\ell$ greater than two. Therefore, if prograde or retrograde g-modes are excited, they should lead to time-varying gravitational potential with angular degrees higher than quadrupole. Because the gravitational potential decays with the $\ell+1$ power of distance from Titan, flybys at low altitudes are most sensitive to time-varying gravitational potentials with $\ell>2$.


\appendix


\section{Equilibrium Tides}\label{app:eq-tides}
\begin{figure}[p!]
\begin{center}
\includegraphics[width=4.2in]{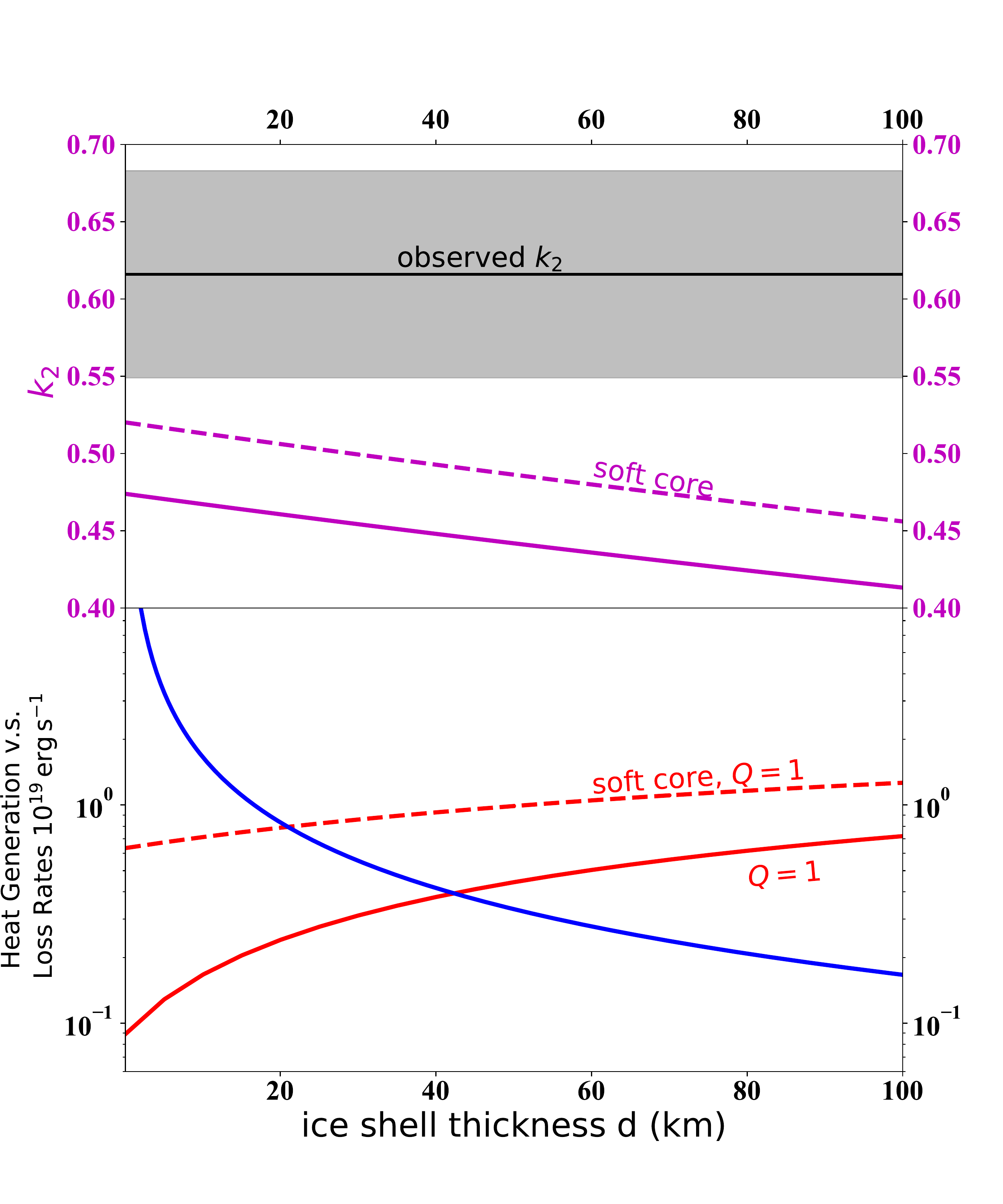}
\caption[Love number and Tidal Heating Rate of Equilibrium Tides of Titan]{
	\label{fig:k2-Heat-eqtides}
	\label{lastfig}	
	This figure shows that equilibrium tides of Titan could not account for the observed $\ktwo$ and generate enough tidal heating to maintain a thin outer ice shell at the same time. {\it The upper panel} shows the theoretical $\ktwo$ due to equilibrium tides of Titan as a function of the thickness of the outermost ice shell, $d$. The solid magenta line adopts the normal values for shear modulus as shown in Table \ref{tab:parameters}. The dashed magenta line denoted by `soft core' artificially scales down the shear modulus of the high-pressure ice layer and the rocky core by a factor of 10. The solid black horizontal line labels the measured $\ktobs=0.616$, and the grey horizontal bar labels the $1\sigma$ uncertainty, $\delta \ktobs=0.067$. {\it The lower panel} shows the maximum tidal heating rate produced by dissipation of elastic energy due to equilibrium tides adopting a tidal quality factor, $Q=1$, by the solid red line. The denotation `soft core' has the same meaning as that in the upper panel. The blue curve shows the conductive heat loss rate, $\loss\sim 4\pi R^2 \kice \DT/d\sim 3.0\times 10^{19}(d/10\km)^{-1}\ergps$. 
	}
\end{center}
\end{figure}

We employ the interior model for Titan with an internal ocean presented in \cite{Rappaport}. In this model, Titan contains a rocky core, a high-pressure ice layer , a water ocean layer, and an outermost ice shell, which are labeled by `c', `hp', `o' and `i1' respectively.  In order to simplify the model, we assume each layer to be incompressible. We adopt all the parameters on the left column of Table~7 in \cite{Rappaport} except for the outer radius for the ocean layer, $\Ro$ and Bulk modulus. We do not fix $\Ro$. Instead, we set it to be $R-d$, where $d$ is the thickness of the outer ice shell which we allow to vary. Bulk modulus is infinite for an incompressible layer. We then use this model to solve for the equilibrium tides through the equilibrium equation \citep{Landau},
\begin{equation}
    \mathbf{\nabla}\cdot\mathbf{\sigma}+\rho\mathbf{g}=0\, ,\label{eq:elastic}
\end{equation}
where $\mathbf{\sigma}$ is the stress tensor and $\mathbf{g}$ is the local gravity acceleration. We apply appropriate boundary conditions, i.e. the Lagrangian perturbation to the $rj$ ($j=r\, ,\, \theta\, ,\, \varphi$) component of the stress tensor must be continuous at each interface. The dynamic quadrupole Love number according to equilibrium tides, $\kteq$, of our model is shown by the solid magenta line in the upper panel of Figure~\ref{fig:k2-Heat-eqtides}. It matches that in Figure~1 of \cite{Rappaport} very well. Note that equilibrium tides have the three components of the quadrupole Love number the same to each other, and thus we denote them by a single number, $\kteq$.

Equilibrium tides yielded by $\Phie$ induces elastic energy, $\Eelaseq$, which is the volumetric integration of the inner product of the stress and strain tensors in the outer ice shell, the high-pressure ice layer and the rocky core. The strain and stress tensors vary periodically with frequency $\norb$, and so does the total elastic energy. If the peak elastic energy, $\left.\Eelaseq\right|_\mathrm{peak}$, damps by a fraction, $1/Q$, with frequency $\norb$, the corresponding tidal heating rate is $\heateq = {\norb\over Q}\left(\left.\Eelaseq\right|_\mathrm{peak}\right)$. Note that the definition of $Q$ in this paper refers to peak energy instead of time-averaged energy. Because $Q$ must not be smaller than unity, the optimal $\left.\heateq\right|_\mathrm{max}$ with $Q=1$ as a function of $d$ is shown by the solid red curve in the lower panel of Figure~\ref{fig:k2-Heat-eqtides}. Also shown in the lower panel is the conductive heat loss rate $\loss$ through the outer ice shell of thickness $d$.

For equilibrium tides, there is a tension between matching $\kteq$ with $\ktobs$ and balancing $\loss$ by $\heateq$. The former needs a thin shell, although $\kteq$ is still more than $1\sigma$ below $\ktobs$ at zero shell thickness. The latter requires a thick shell because the conductive heat loss rate $\loss$ becomes smaller for a thicker shell and also a thicker shell means more volume to store elastic energy which dissipates into heat. Unfortunately for Titan, this tension could not be resolved by simply adjusting the shell thickness. 

This tension is quantitatively illustrated by Figure~\ref{fig:k2-Heat-eqtides}. Equating the optimal $\left.\heateq\right|_\mathrm{max}$ with $\loss$ leads to $d\sim 42\km$. The corresponding $\kteq\sim 0.45$, which is $2\sigma$ below $\ktobs$. A reasonable $Q\geq 10$ requires $d\geq 165\km$, making $\kteq\leq 0.385$, more than $3\sigma$ below the observed Love number. One may artificially tune the shear modulus for the rocky core and the high-pressure ice layer to much smaller values. This would enlarge $\kteq$ as well as $\heateq$. One does not want to tune the shear modulus of the outer ice shell to a smaller value because that would enhance $\kteq$ only by a tiny amount but make $\heateq$ smaller. We make $\muhp$ and $\muc$ ten times smaller and see that this tuning makes the problem less serious but does not really resolve it, as demonstrated by dashed curves in Figure~\ref{fig:k2-Heat-eqtides}. We must admit that we do not explore artificial tuning exhaustively, and therefore we cannot claim confidently that there does not exist a set of artificially-tuned parameters that makes equilibrium tides work.

\section{Dynamic Quadrupole Love Number}\label{app:love-number-3-components}
The time-varying quadrupole tidal potential exerted on Titan by Saturn, to the leading order of $e$, is \citep{MD}
\begin{eqnarray}
\Phie(r,\theta,\varphi,t) =&& {3\over 2} e \norb^2 r^2 \Ptz(\cos\theta)\cos(\norb t)\,\label{eq:Phiesym}\\
&& -{7\over 8} e \norb^2 r^2\Ptt(\cos\theta) \cos(2\varphi- \norb t)\,\label{eq:Phiepro}\\
&& +{1\over 8} e \norb^2 r^2 \Ptt(\cos\theta)\cos(2\varphi+\norb t)\, .\label{eq:Phieretro}
\end{eqnarray}
Here $\theta$ and $\varphi$ are the colatitude and longitude of Titan measured in the frame rotating at Titan's spin rate, $\Omega_\mathrm{T}$. Titan is synchronized with its orbit, and thus $\Omega_\mathrm{T}=\norb$.
The first term (Equation~\ref{eq:Phiesym}), denoted by $\Phiesym$, is axisymmetric because it is independent of $\varphi$. The second term (Equation~\ref{eq:Phiepro}), denoted by $\Phiepro$, is prograde since its azimuthal phase speed is positive. The last term  (Equation~\ref{eq:Phieretro}), denoted by $\Phieretro$, is retrograde.

The time-varying quadrupole potential induced by the tidal deformation of Titan in response to $\Phie$ may be parameterized as
\begin{equation}
    \Phieself(r\geq R,\theta,\varphi,t)=\left(R\over r\right)^3\left[\ksym\Phiesym+\kpro\Phiepro+\kretro\Phieretro\right]_{r=R}\,  ,\label{eq:Phieself}
\end{equation}
where $\ksym$, $\kpro$ and $\kretro$ are the axisymmetric, prograde and retrograde components of the dynamic quadrupole Love number. 

\section{Relation between the radial displacements at $R$ and at $\Ro$}\label{app:xirsurf-xiro-relation}

We assume the outermost ice shell above the internal ocean is incompressible. An ocean wave with frequency $\norb$ deforms the shell from below. Since the sound travel time through the ice shell is much shorter than the period of the ocean wave. The deformation of the ice shell reacts instantaneously to the ocean wave. Therefore, the equilibrium equation~(\ref{eq:elastic}) applies. The outer boundary of the ice shell at $r=R$ is a free surface, and thus the Lagrangian perturbation to the stress tensor must vanish,
\begin{equation}
\mathbf{\delta\sigma}=\mathbf{0}\, .
\end{equation}
The lower boundary of the ice shell is in touch with a shear-free ocean, and thus at $r=\Ro$,
\begin{equation}
\delta\sigma_{rj} = 0\, , (j=\theta,\varphi)\, .
\end{equation}
The radial displacement must be continuous at $r=\Ro$. Applying these boundary conditions and solving the equilibrium equation, we get a relation between the radial displacement at the surface and that at $r=\Ro$. If the angular shape of the radial displacement is $\Ptz(\cos\theta)$, then the relation is approximately
\begin{equation}
    \left({\xirsurf\over \left.\xir\right|_{r=\Ro} }\right)_\mathrm{20} \approx 1-d\left({4\over 11 R}+{6g\rhoi\over 55\mui}\right) \approx 1-0.0018\left(d\over 10\km\right)\, ,
\end{equation}
to the leading order of $d$. For a different angular shape, the integers in the bracket after the first $\approx$ would change, but it is still true that $\xi_r^\mathrm{surf}$ is well approximated by $\left.\xir\right|_{r=\Ro}$. If there is a g-mode excited resonantly and its first half-wavelength has a radial displacement $\left.\xir\right|_{r=\Ro}$. This mode would force the same radial displacement for the ice shell.


The horizontal displacement at the outer surface of the ice shell takes the format of $\xihsurf\, R\, {\bf \nabla}(\Ptz(\cos\theta))$.
We get
\begin{equation}\label{eq:xihsurf-xiro-relation}
 \left({\xihsurf\over \left.\xir\right|_{r=\Ro} }\right)_\mathrm{20} \approx {3\over 11}-{R g\rhoi\over 55\mui}-d\left({56\over 121 R}{14 g\rhoi\over 605\mui}\right)\approx 0.255-1.88\times 10^{-3}\left(d\over 10\km\right)\, .
\end{equation}

\section{Tidal Heating Scenarios other than Equilibrium Tides}\label{app:tidal-heating}
If tidal heating could keep a very thin outer ice shell, e.g. $d\sim 10\km$, then the dynamic Love number due to equilibrium tides reaches its maximum value. Although $\kteqmax$ is still smaller than $\ktobs$, at least the tension between $\kteq$ and $\ktobs$ could be partly reconciled. This is also an important question for understanding Titan's orbital dynamics. Because tidal heat originates from the epicyclic energy of Titan, tidal heating leads to damping of the orbital eccentricity. 

We have shown in Appendix~\ref{app:eq-tides} that tidal heating rate due to dissipation of elastic energy in the solid parts of Titan is too small. We propose that g-modes in the presumably stably-stratified ocean may be resonantly excited by tides. Then it is worth checking whether damping of ocean waves can lead to any significant tidal heating. The answer is no. We briefly list our order-of-magnitude estimates below for interested readers. 

\subsection{Heating due to damping a resonant g-mode}
Because g-mode mainly moves horizontally, $\xih/\xir\sim N/\norb\gg 1$, a g-mode could potentially contain a lot of kinetic energy. This feature distinguishes g-modes from equilibrium tides. For the latter, elastic energy greatly exceeds kinetic energy. However, stable stratification slows down the group velocity of ocean wave, and thus it takes a long time for a g-mode to grow. In order to store a lot of kinetic energy in a g-mode, one must give it enough time to grow. During this time, one cannot damp the mode significantly, otherwise the mode's growth should have stopped. Although a g-mode can contain a lot of kinetic energy, this advantage is traded off by its slow damping. Therefore, tidal heating due to damping of a g-mode is small. 

The kinetic energy of a resonant g-mode is roughly
\begin{equation}
    \Emode\sim 4\pi R^2 \rhoo H (\norb\, \xih^\mathrm{mode})^2\, ,
\end{equation}
where $\rhoo$ is the density of water. Adopting Equations~(\ref{eq:wave-eq}), (\ref{eq:eqr-eqh}), (\ref{eq:ttra}), (\ref{eq:tcoh}) and (\ref{eq:mode-wave}), we get
\begin{equation}
    \xih^\mathrm{mode}\sim \left.\xireq\right|_\mathrm{surf}{\pi R\over\sqrt{\Lambda}H}{\Dw\over \sqrt{\dw^2+\gamma^2}}\, .
\end{equation}
The tidal heating rate due to dissipation of $\Emode$ is maximized when $\dw=\gamma$ because $\dw^2+\gamma^2\geq 2\dw\gamma$. We obtain
\begin{eqnarray}
    \dot E_\mathrm{mode}^\mathrm{max}&\sim & (\gamma \Emode)_\mathrm{max} \sim 2\pi^3  (\norb\left.\xireq\right|_\mathrm{surf})^2\norb {R^4 \rhoo\over\Lambda H} {\Dw^2\over \omega\gamma}\\
    &\sim& 3.0\times 10^{16}\left[H\over 300\km\right]^{-1}{\Dw^2\over\omega\gamma}\,\mathrm{erg\, s^{-1}}\, ,
\end{eqnarray}
where we have applied $\left.\xireq\right|_\mathrm{surf}\sim 460\,\mathrm{cm}$ which is the maximum radial displacement due to equilibrium tides when $d=0\km$. 
Meanwhile, the mode's radial displacement is
\begin{equation}
    \xir^\mathrm{mode}\sim \left.\xireq\right|_\mathrm{surf} {\Dw^2\over\sqrt{2}\gamma\omega}\, .
\end{equation}
Note that the maximal tidal heating rate and the mode's radial displacement have a common factor $\Dw^2/(\gamma\omega)$ in their expressions. The observed dynamic Love number constrains the total surface radial displacement of Titan. Therefore the factor $\xir^\mathrm{mode}/\left.\xireq\right|_\mathrm{surf}$ cannot exceed $\ktobs/\kteqmax\approx 1.3$. It follows that $\Dw^2/(\gamma\omega)<1.8$. Thus we get
\begin{equation}
    \dot E_\mathrm{mode}^\mathrm{max} < 5.4\times 10^{18}\left[H\over 300\km\right]^{-1}\ergps\, ,
\end{equation}
which is still one order of magnitude smaller than $\loss\sim 3.0\times 10^{19}(d/10\km)^{-1}\ergps$. 

One might say that if the depth of the ocean is $30\km$, then the optimal tidal heating rate due to a g-mode could be big enough to match $\loss$. But in order to fulfill this optimal tidal heating, one has to come up with a physical reason for $\dw=\gamma$, i.e. the tidal frequency $\norb$ falls particularly close to a g-mode. Normally we expect the frequency mismatch $\dw$ is similar to frequency separation of neighboring g-modes, $\Dw$, which is much larger than the linewidth of a g-mode, $\gamma$. We find the optimal case implausible although it cannot be ruled out. 

\subsection{Heating due to resonance ocean wave in a very thin ocean}
As the ocean becomes very thin, the dispersion relation for ocean waves becomes
\begin{equation}
    \omega^2 = g\kh^2 H\, .
\end{equation}
Ocean wave is a thin ocean is evanescent radially. It travels horizontally at the group velocity 
\begin{equation}
    v_g\sim {\partial\omega\over\partial \kh} \sim \sqrt{g H}\, .
\end{equation}
The kinetic energy in the ocean wave is
\begin{equation}
    E_\mathrm{wave}\sim 4\pi R^2 \rhoo H (\norb\xih^\mathrm{wave})^2 \sim 4\pi (\norb\left.\xireq\right|_\mathrm{surf})^2 {R^4\rhoo\over \Lambda H}\, ,
\end{equation}
wich we have used Equations~(\ref{eq:wave-eq}) and (\ref{eq:eqr-eqh}). The tidal heating rate is maximized if the ocean wave damps after it travels to the other side of Titan in time $\sim \pi R/v_g\sim \pi R/\sqrt{g H}$. Therefore, the maximal heating rate is
\begin{equation}
    \dot E_\mathrm{wave}^\mathrm{max}\sim {E_\mathrm{wave}\over \pi R/\sqrt{g H}} \sim {4R^3\rhoo\over \Lambda}\left(g\over H\right)^{1/2}(\norb\left.\xireq\right|_\mathrm{surf})^2\, . 
\end{equation}
In order for $\dot E_\mathrm{wave}^\mathrm{max}$ to balance $\loss\sim 3\times 10^{19}(d/10\km)^{-1}\ergps$, one requires
\begin{equation}
    H\sim 3.8\times 10^{-3}\left(d\over 10\km\right)^2\km\, .
\end{equation}
Such a thin ocean seems very unlikely.

\subsection{Inertial Wave}
Since Titan is rotating, there exists inertial waves (IW) in the ocean. The restoring force for inertial waves is Coriolis force. According to \cite{Ogilvie-2013}, with $\alpha=\Rc/R\approx 0.7$, we read from their Figure~4. that the imaginary Love number is at most $2.4\times 10^{-5}$ at $\omega/\Omega_T=\pm 1$ with the $Y_2^0$ component. The corresponding eccentricity damping timescale \citep{Peale-review}
\begin{equation}
    \taue={2\over 21}{\MT\over\MS}\left(a\over R\right)^5{1\over \mathrm{Im}(k_2)}{1\over\norb}\, ,
\end{equation}
leads to a tidal heating rate
\begin{equation}
    \dot E_\mathrm{IW}\sim {\MT(a\norb e)^2\over\taue}\sim 7\times 10^{15}\ergps\, .
\end{equation}
This is too small to maintain the outer ice shell at a thickness much smallter than $100\km$.

\ack
We are grateful to Daniele Durante and Luciano Iess who generously shared their latest observational results. We thank Peter Goldreich, Jeremey Goodman and Dong Lai for their insightful comments and suggestions. Jing Luan is supported at the Institute for Advanced Study.


\label{lastpage}


\bibliography{bibliography.bib}

\bibliographystyle{plainnat}


\clearpage	


\begin{table}
\begin{center}
\textbf{Parameters and Symbols}
\begin{tabular}{c|l||c|l}       
\hline
\hline
$\MT$ & $1.34\times 10^{23}\kg$, mass of Titan & $\rhoi$ & $0.92\,\mathrm{g\, cm^{-3}}$\\
$R$ & $2575\km$, radius of Titan & $\rhoo$ & $1.0\,\mathrm{g\, cm^{-3}}$\\
$d$ & thickness of the outermost ice shell &$\rhohp$ & $1.31\,\mathrm{g\, cm^{-3}}$ \\ 
$I$ & $0.35^\circ$, orbit inclination with respect to & $\rhoave$ & $1.881\,\mathrm{g\, cm^{-3}}$ \\
 & the equatorial plane of Saturn &$\mui$ & $3.3\,\mathrm{GPa}$ \\
$e$ & orbital eccentricity, current value $\enow\approx 0.0288$ &$\muo$ & $0.0\,\mathrm{GPa}$ \\
$\norb$ & $4.56\times 10^{-6}\,\mathrm{rad\, s^{-1}}$, orbit mean motion &$\muhp$ & $4.6\,\mathrm{GPa}$ \\
$a$ & $1.22\times 10^{6}\km$, orbit semi-major axis & $\muc$ & $60.0\,\mathrm{GPa}$\\
$\ktobs$ & $0.616\pm 0.067$, measured quadrupole Love number of Titan & $\Rhp$ & $2275\km$ \\
$\kice $ & $2\times 10^5\,\mathrm{erg\, K^{-1}\, cm^{-1}\, s^{-1}}$,  thermal conductivity of ice &$\Rc$ & $1670\km$ \\
$\cp$ & $2\,\mathrm{kJ\, kg^{-1}\, K^{-1}}$, specific heat capacity of ice & $\Ro$& $R-d$  \\
$\DT$ &$180\K$, temperature contrast through the ice shell\\
$g$ & $135\,\mathrm{cm\, s^{-2}}$, surface gravity of Titan\\
\hline
\end{tabular}
\caption[Parameters and Symbols]
	{\label{tab:parameters}	
	\label{lasttable}
	The right panel quotes Table~7 of \cite{Rappaport}. The subscripts `i1', `o', `hp' and `c' label the outer most ice shell, the liquid ocean layer, the high-pressure ice layer and the rocky core respectively. $\Ro$ refers to the outer radius of the ocean layer, etc. $\rho$ denotes density and $\mu$ shear modulus. As the thickness of the outer ice shell is varied, the density of the core is adjusted to keep the same average density, $\rhoave$. 
	}
\end{center}
\end{table}

\clearpage


\end{document}